\documentclass[twocolumn,floatfix,prb,showpacs,10pt,aps]{revtex4-1}
\usepackage[latin1]{inputenc}
\usepackage[T1]{fontenc}
\usepackage{graphicx}
\usepackage{amssymb}
\usepackage{amsmath}
\usepackage{color}
\usepackage{psfrag}
\usepackage{epsfig}
\usepackage{bbm}
\usepackage{bm}
\usepackage{dsfont} 
\usepackage{hyperref}

\newcommand{\ket}[1]{| #1 \rangle}

\newcommand{\rb}[1]{\left( #1 \right)}

\newcommand{\ew}[1]{\langle #1 \rangle}
\newcommand{\beq}{\begin{eqnarray}}
\newcommand{\eeq}{\end{eqnarray}}

\newcommand{\op}[2]{| #1 \rangle \langle #2 |}

\newcommand{\eq}[1]{Eq.~(\ref{#1})}
\newcommand{\fig}[1]{Fig.~\ref{#1}}

\newcommand{\trace}[1]{\mathrm{Tr}\left\{#1\right\}}

\newcommand{\bld}[1]{\boldsymbol #1}

\newcommand{\eww}[1]{\langle\! \langle #1\rangle\! \rangle}

\begin{document}
\title{Bunching and anti-bunching in electronic transport}
\author{
Clive Emary,
Christina P\"oltl,
Alexander Carmele, 
Julia Kabuss,
Andreas Knorr,
and 
Tobias Brandes 
}
\affiliation{
  Institut f\"ur Theoretische Physik,
  Hardenbergstr. 36,
  TU Berlin,
  D-10623 Berlin,
  Germany
}
\date{\today}
\begin{abstract}
In quantum optics the $g^{(2)}$-function is a standard tool to investigate photon emission statistics.
We define a $g^{(2)}$-function for electronic transport and use it to investigate the bunching and anti-bunching of electron currents.  
Importantly, we show that super-Poissonian electron statistics do not necessarily imply electron bunching, and that sub-Poissonian statistics do not imply anti-bunching. 
We discuss the information contained in $g^{(2)}(\tau)$ for several typical examples of transport through nano-structures such as few-level quantum dots. 
\end{abstract}
\maketitle
%%%%%%%%%%%%%%%%%%%%%%%%%%%%%%%%%%%%%%%%%%%%%%%%%%%%%%%%%%%%%%%%%%%%%%%%%

\begin{figure}[tb]
  \begin{center}
\includegraphics[width=\columnwidth,clip]{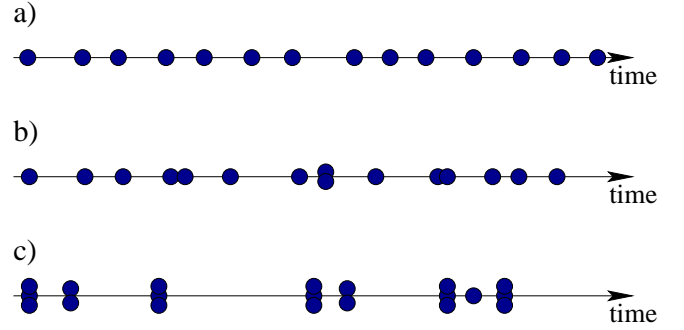}
  \caption{ 
    Sketch of a) anti-bunched, b) Poissonian and c) bunched photon (or electron) emission events. 
    \label{bunches}
 }
  \end{center}
\end{figure}

Current noise has long-since been established as an important tool for studying the physics of transport through mesoscopic and nano-scale conductors \cite{FCS1,el-noise1,el-noise2,FCS2,FCS3}.
The character of the noise is typically assessed by considering the Fano factor, the ratio of the zero-frequency noise to the current \cite{el-noise2}, and comparing with a Poisson process for which the Fano factor is equal to one.  Systems with $F<1$ are described as sub-Poissonian (non-interacting systems fall in this class \cite{el-noise1}) and systems which have $F>1$ are called super-Poissonian.
A common interpretation of this comparison is that a super-Poissonian Fano factor indicates a {\em bunching} of the current's constituent electrons, whereas sub-Poissonian values indicates {\em anti-bunching} (Fig.~\ref{bunches} ).

In this paper we directly investigate bunching and anti-bunching in electronic transport as a phenomenon in the time domain through the introduction of a second-order correlation function $g^{(2)}(\tau)$, analogous to that used in quantum optics \cite{ScullyBook,Loudon-book,Mandel-Wolf}.  Within a  quantum master equation (QME) framework in the appropriate limit, the $g^{(2)}$-function is seen to be proportional to the conditional probability that, given an electron is emitted into the collector at time $t=0$, a further such jump is observed a time $\tau$ later.   
Following quantum optics, we identify
\begin{align}
\begin{array}{cc}
  g^{(2)}(0) ~>~ g^{(2)}(\tau) & \mathrm{bunching}\\
  g^{(2)}(0) ~<~ g^{(2)}(\tau) & \mathrm{anti-bunching}
\end{array}
\label{BvsAB}
\end{align}
since bunching means that particles are more likely to be emitted together than apart, and conversely for anti-bunching.
By relating our $g^{(2)}$-function to the correlation function between the current at two different times, we clarify the relationship between the $g^{(2)}$-function, 
(anti-) bunching and the Fano factor.

%%%%%%%%%%%%%%%%%%%%%%%%%%%%%%%%%%%%%%%%%%%%%%%%%%%%%%%%%%%%%%%%%
%%%%%%%%%%%%%%%%%%%%%%%%%%%%%%%%%%%%%%%%%%%%%%%%%%%%%%%%%%%%%%%%%

We then investigate bunching and anti-bunching in several widely-discussed transport models in the Coulomb blockade (CB) regime (see \fig{models}). 
This analysis shows that the simple picture relating super-Poissonian Fano factors to bunching and sub-Poissonian ones to anti-bunching is often an oversimplification, and can even be outright wrong. 
In particular we discuss a simple quantum-dot (QD) model which has a Fano factor less than one, and is thus sub-Poissonian, and yet has $g^{(2)}(0) ~>~ g^{(2)}(\tau)$ for all $\tau>0$ such that, according to \eq{BvsAB}, the electron- flow is completely bunched.  We also give a model for which the converse is true, i.e. we find a super-Poissonian Fano factor in conjunction with electron anti-bunching.
These results mirror the work of Singh \cite{anti-super} and Zou and Mandel \cite{bunch-sub}, who have made similar points for quantum-optical systems.

This paper proceeds by first reviewing the situation in quantum optics.  We then define by analogy our $g^{(2)}$-function for transport systems and examine its properties, including its relationship with current fluctuations and the zero-frequency Fano factor. We then consider our concrete examples and conclude.

%%%%%%%%%%%%%%%%%%%%%%%%%%%%%%%%%%%%%%%%%%%%%%%%%%%%%%%%%%%%%%
%%%%%%%%%%%%%%%%%%%%%%       OPTICS       %%%%%%%%%%%%%%%%%%%%
%%%%%%%%%%%%%%%%%%%%%%%%%%%%%%%%%%%%%%%%%%%%%%%%%%%%%%%%%%%%%%
\section{The second-order correlation function in quantum optics}

The second-order degree of coherence of the electric field at position $\mathbf{r}$ and times $t$ and $t+\tau$; $\tau \ge 0$ is defined as \cite{ScullyBook}
\begin{align}  
  \label{g2full}
  g^{(2)}(\mathbf{r},t,\tau) \equiv \hspace{5.7cm} \nonumber \\
  \frac{
    \ew{
      E^{(-)}(\mathbf{r},t) 
      E^{(-)}(\mathbf{r},t+\tau) 
      E^{(+)}(\mathbf{r},t+\tau) 
      E^{(+)}(\mathbf{r},t) 
    }
  }
  {
    \ew{
      E^{(-)}(\mathbf{r},t) 
      E^{(+)}(\mathbf{r},t) 
    }
    \ew{
      E^{(-)}(\mathbf{r},t+\tau) 
      E^{(+)}(\mathbf{r},t+\tau)
    }  
  }
  .
\end{align}
Here, linear polarisation in direction $ \mathbf{e}$ is assumed such that $ E^{(\pm)} \equiv \mathbf{e}\cdot\mathbf{E}^{(\pm)} $.  
In the stationary limit and suppressing the position dependence, we have 
\begin{align}
  \label{g2traditional}
  g^{(2)}(\tau) \equiv
  \frac{
    \ew{
      E^{(-)}
      E^{(-)}(\tau) 
      E^{(+)}(\tau) 
      E^{(+)}
    }
  }
  {
    \ew{
      E^{(-)}
      E^{(+)}
    }^2
  }
  ,
\end{align}
with $E^{(\pm)}=E^{(\pm)}(\mathbf{r},t=0)$.

The next step is a re-formulation of  the traditional Heisenberg operator definition \eq{g2traditional} in the (somewhat more flexible)
Schrödinger (master equation) picture. 
For example, in the theory of resonance fluorescence \cite{Carmichael2003}, one can express the electric field in the far-field limit in terms of the atomic (two-level) operators
$\sigma_\pm(t)$  via
\begin{align}
  \mathbf{E}^{(+)}(\mathbf{r},t) = C(\mathbf{r} -\mathbf{r}_0)
  \mathbf{e}
  \sigma_-(t-t_0)
  ,
\end{align}
where $C(\mathbf{r})$ is some position-dependent constant, $ \mathbf{e}$ the appropriate polarisation vector and $t_0 = |\mathbf{r}-\mathbf{r}_0|/c$ the time-of-flight from atom to detector  \cite{ScullyBook}.
With this substitution, the stationary $g^{(2)}$-function of the field can be written purely in terms of atomic degrees of freedom as
\begin{align}
  g^{(2)}(\tau) =
  \frac{
    \ew{
      \sigma_+
      \sigma_+(\tau) 
      \sigma_-(\tau) 
      \sigma_-
    }
  }
  {
    \ew{
      \sigma_+
      \sigma_-
    }^2
  }
  \label{g2sigmas}
  ,
\end{align}
where all the constants $C$ cancel. The main technical step is now a formulation of the dissipative 
quantum dynamics in terms of a quantum master equation with Liouvillian $\mathcal{W}$, $\dot \rho = \mathcal{W}  \rho$, for the state  $\rho$ of the atom in Born-Markov approximation. 
The quantum regression theorem then
allows one to write \cite{Carmichael2003}
\begin{align}
  g^{(2)}(\tau) =
  \frac{
    \trace{
      \sigma_- 
      e^{ \mathcal{W} \tau}[
      \sigma_-
      \rho_\mathrm{stat}
      \sigma_+]
      \sigma_+ 
    }
  }
  {
    \trace{
      \sigma_-
      \rho_\mathrm{stat}
      \sigma_+
    }^2
  }
  ,
  \label{QOX:g2_def_sigs}
\end{align}
where $\rho_\mathrm{stat}=\rho(t\to\infty)$ is the stationary state of the atom at large times $t$. 
One then introduces the downwards-jump super-operator $\mathcal{J}_\gamma$ through its action on an arbitrary density matrix $\rho$
\begin{align}
  \mathcal{J}_\gamma
%(\tau) 
\rho=\gamma  \sigma_-
%(\tau)
\rho \sigma_+
%(\tau)
  ,
\label{ph-jump}
\end{align}
where $\gamma$ is the spontaneous emission rate, leading to 
\begin{align}
  g^{(2)}(\tau)
  =&
  \frac{
    \trace{
      \mathcal{J}_\gamma  e^{ \mathcal{W} \tau}
      \mathcal{J}_\gamma
      \rho_\mathrm{stat}
    }
  }
  {
    \trace{
      \mathcal{J}_\gamma
      \rho_\mathrm{stat}
    }^2
  }.
  \label{G2QO:JOmJ}
\end{align}
This equation makes it clear that, at least from a resonance fluorescence perspective, the $g^{(2)}$ function measures the correlation between two system jumps separated by a time $\tau$.

%%%%%%%%%%%%%%%%%%%%%%%%%%%%%%%%%%%%%%%%%%%%%%%%%%%%%%%%%%%%%%
%%%%%%%%%%%%%%%%%%%%%%      TRANSPORT     %%%%%%%%%%%%%%%%%%%%
%%%%%%%%%%%%%%%%%%%%%%%%%%%%%%%%%%%%%%%%%%%%%%%%%%%%%%%%%%%%%%
\section{A second-order correlation function for quantum transport \label{sec:trans}}

We consider here transport systems in the infinite-bias limit such that the time-evolution of the system density matrix can be described by a Markovian master equation 
$
  \dot \rho = \mathcal{W}  \rho
$ 
\cite{FCS3}. We assume that we are only interested in the current in a single lead, assumed to be a collector, and decompose the Liouvillian as $  \mathcal{W} = \mathcal{W}_0 + \mathcal{J} $ where, similarly to \eq{ph-jump}, the jump super-operator $\mathcal{J}$ describes incoherent transitions of the system.  Here $\mathcal{J} $ describes  the emission of an electron into the lead in question.  The remaining part of the Liouvillian, $\mathcal{W}_0$, describes the evolution of the system without such jumps (jumps to and from other leads are included in $\mathcal{W}_0$).
When we discuss our examples in the next section, we will give the kernels in the full-counting-statistics form  $\mathcal{W}(\chi)=\mathcal{W}_0 + \mathcal{J}e^{i \chi}$ with $\chi$ the counting field such that $\mathcal{W} = \mathcal{W}(\chi=0)$ and $\mathcal{J}=-i \frac{d}{d\chi} \mathcal{W}(\chi)|_{\chi=0}$.

We then define the $g^{(2)}$-function for transport in analogy with \eq{G2QO:JOmJ}, replacing the optical jump superoperators with  their transport counterparts:
\begin{align}
  g^{(2)}(\tau)
  \equiv
  \frac{
    \eww{
      \mathcal{J}\Omega(\tau) 
      \mathcal{J}
    }
  }
  {
    \eww{
      \mathcal{J}
    }^2
  }
  \label{g2trans}
  ,
\end{align}
where we have explicitly included the master equation propagator $\Omega(\tau) = e^{\mathcal{W}\tau}$, and $\eww{...}$ denotes the stationary expectation value,  $\eww{A} = \trace{A \rho_\mathrm{stat}}$ with $\mathcal{W}\rho_\mathrm{stat}=0$, $\rho_\mathrm{stat}$ assumed unique. 
The  interpretation of $g^{(2)}(\tau)$ is analogous to the quantum optical case, and in particular the designation of bunching and anti-bunching follows \eq{BvsAB}.

Whilst we have defined our transport $g^{(2)}$-function here by analogy between master-equation formalisms (i.e. in terms of jump operators), result \eq{g2trans} can also be derived from a microscopic Heisenberg-picture expression similar to \eq{g2full}. This derivation is discussed in the appendix.

\subsection{Properties}

We use a Liouville space ket $|0 \rangle\rangle=\rho_\mathrm{stat} $  
and bra $\langle\langle \tilde{0}|$ such that the trace over the stationary state is written $\eww{A} = \langle\langle \tilde{0}|A |0 \rangle\rangle$. 
It is then useful to define the projector onto the stationary state $\mathcal{P}=   |0 \rangle\rangle    \langle\langle \tilde{0}| $ via $ \mathcal{W}\mathcal{P} =  \mathcal{P}\mathcal{W} =0 $, along with its compliment $ \mathcal{Q} = \mathds{1} - \mathcal{P}$. This allows us to decompose the propagator as $ \Omega(\tau) = \mathcal{P} +  \mathcal{R}(\tau)$ with irreducible part
$
   \mathcal{R}(\tau) = \mathcal{Q} \Omega(\tau) \mathcal{Q}
$.
Using $\eww{ \mathcal{J}\mathcal{P} \mathcal{J}}=  \langle\langle \tilde{0}| \mathcal{J}\mathcal{P} \mathcal{J} |0 \rangle\rangle=\eww{ \mathcal{J}}^2$,  we find
\begin{align}
  g^{(2)}(\tau)
  =
  1+
  \frac{
    \eww{
      \mathcal{J}
      \mathcal{R}(\tau) 
      \mathcal{J}
    }
  }
  {
    \eww{
      \mathcal{J}
    }^2
  }
  ;\quad 
  \tau >0
  \label{G2QT:1+JRJ}
  .
\end{align}
In this long-time limit, all non-zero eigenvalues of $\mathcal{W}$ have negative real-parts and $\mathcal{R}(\tau)$ decays with $\tau\to \infty$ such that
$
  g^{(2)}(\tau \to \infty) = 1
$.

In the limit of strong Coulomb blockade such that at most one excess electron occupies the system at any time, we have $\mathcal{J}^2 = 0$, since the exit of an electron from the system leaves it empty and a second jump can not take place immediately. In this case
\begin{align}
  g^{(2)}(\tau = 0) =0
  .
\end{align}
This leads to the conclusion that, according to the definition \eq{BvsAB}, a system in the strong CB regime is always anti-bunched, since $g^{(2)}(\tau = 0) <g^{(2)}(\tau )$ as $g^{(2)}(\tau )$ is always positive. 

%%%%%%%%%%%%%%%%%%%%%%%%%%%%%%%%%%%%%%%%%%%%%%%%%%%%%%%%%%%%%%%%%%%%%
%\subsection{Fano factor}

The current noise at finite frequency is defined as $ S (\omega) \equiv \frac{1}{2} \int dt e^{i \omega t} \ew{\left\{\delta I(t),\delta I(0)\right\}}$.  In the QME approach with unidirectional transport, this can be written in terms of system quantities as \cite{FCS3}
\begin{align}
  S(\omega) = \eww{\mathcal{J}} 
  + \eww{\mathcal{J}\left[
    \mathcal{R}(-i\omega) + \mathcal{R}(i\omega) 
  \right]\mathcal{J}}
  , 
\end{align}
with Laplace-transformed irreducible propagator $ \mathcal{R}(z) = \left[z-\mathcal{QWQ}\right]^{-1}$.  The mean current reads $\ew{I} = \eww{\mathcal{J}}$.
Transformation back into the time-domain yields \cite{HDH93,KSW04}
\begin{align}
  S(\tau) =\ew{I} \delta(\tau)
  + \eww{\mathcal{J} \mathcal{R}(|\tau|)\mathcal{J}}
  ;\quad \forall \tau
  . 
\end{align}
Thus providing we set $\tau > 0 $, we can equate 
\begin{align}
   g^{(2)}(\tau) = \frac{S(\tau)}{\ew{I}^2} +1 
   ;\quad 
   \tau > 0
   . \label{S-tau}
\end{align}
Or, in the frequency domain, we have
\begin{align}
  S(\omega) \equiv& \int_{-\infty}^{+\infty} d \tau e^{i \omega t} S(\tau)
  \nonumber\\
  =&\ew{I}
  + 2 \ew{I}^2 
    \int_{0^+}^{\infty} d \tau
    e^{i \omega \tau}
    \rb{
      g^{(2)}(\tau)
      -1
    }
    .
\end{align}
These results show that, in the infinite-bias limit, the $g^{(2)}$-function can be directly related to the current-current correlation function either in frequency or time domain.

Defining the Fano factor as the ratio of zero-frequency noise to the mean current, we have
%\begin{align}
\beq
  \label{Fano}
   F(0) \equiv
   \frac{S(\omega=0)}{\ew{I}} 
   %\nonumber\\
   %=& 
   =
   1
   + 2  \ew{I}
    \int_{0^+}^{\infty} d \tau
    \rb{
      g^{(2)}(\tau)
      -1
    }.
\eeq
%\end{align}
This result for the Fano factor coincides with the optical Mandel $Q$ factor $Q=F(0)-1$ in the limit of long counting times $T\rightarrow\infty$.\cite{foot2}

% waiting times
We also note that the $g^{(2)}$-function defined here bears some resemblance to the waiting-time distribution \cite{waiting1,waiting2}
 \begin{align}
   w(\tau)
   =
   \frac{
     \eww{
       \mathcal{J} e^{\mathcal{W}_0 \tau}
       \mathcal{J}
     }
   }
   {
     \eww{
       \mathcal{J}
     }
   },
   \quad 
   \tau >0
   ,
 \end{align}
 but there are two differences: (i) the normalisation is different, and (ii) the waiting-time propagator is defined with Liouvillian $\mathcal{W}_0$ which excludes additional jumps within the interval $\tau$.  The $g^{(2)}$-function, in contrast, contains the full propagator $\mathcal{W}$.

%%%%%%%%%%%%%%%%%%%%%%%%%%%%%%%%%%%%%%%%%%%%%%%%%%%%%%%%%%%%%%
%%%%%%%%%%%%%%%%%%%%%%      EXAMPLES      %%%%%%%%%%%%%%%%%%%%
%%%%%%%%%%%%%%%%%%%%%%%%%%%%%%%%%%%%%%%%%%%%%%%%%%%%%%%%%%%%%%
\section{Examples}
%%%%%%%%%%%%%%%%%%%%%%%%%%%%%%%%%%%%%%%%%%%%%%%%%%%%%%%%%%%%%%%%%
\begin{figure}[tb]
  \begin{center}
\includegraphics[width=\columnwidth,clip]{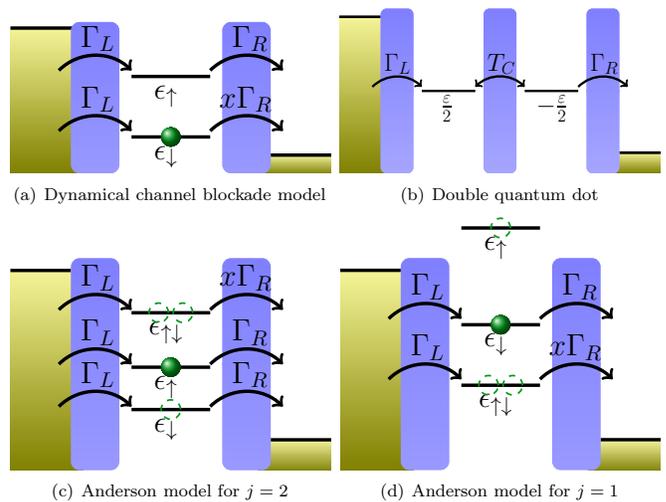}
  \caption{
    Sketch of four of the five transport models discussed here:
    (a) The dynamical channel blockade model;
    (b) the double quantum dot;
    (c) the Anderson model ($j=2$); and 
    (d) the Anderson model with only one singly-occupied level participating in transport ($j=1$). 
    In each case, electrons tunnel with the rate $\Gamma_L$ ($\Gamma_R$) into (out of) the quantum dot systems except when the outgoing rate is modified by the factor $x$.
    \label{models}
  }
  \end{center}
\end{figure}
%%%%%%%%%%%%%%%%%%%%%%%%%%%%%%%%%%%%%%%%%%%%%%%%%%%%%%%%%%%%%%%%%

We now discuss the features of the transport $g^{(2)}$-function for several master-equation models, widely used to describe the electronic transport through CB systems such as quantum dots (QDs) or molecules.  
In each case, we consider unidirectional transport in a two lead set-up, where electrons tunnel into the system from the left lead with the rate $\Gamma_L$, and out of the system into the right with rate $\Gamma_R$.

\subsection{Single resonant level}
Our first example, the single resonant level (SRL), describes the transport through a single level (in a QD, for example) in the strong CB regime. Written in a basis such that the relevant part of the density matrix is collected into the vector $\bld{\rho}=\{\rho_0,\rho_1\}$, where 0 and 1 denote an empty or occupied level, the Liouvillian reads
\begin{align}
  \mathcal{W}^\mathrm{SRL}(\chi) = 
  \rb{
    \begin{array}{cc}
     - \Gamma_L 
         & \Gamma_R   e^{i \chi}  \\
      \Gamma_L & -\Gamma_R  \\
    \end{array}    
  }
  \label{QD_kernel}
  ,
\end{align}
The  steady state current is given by
$ \ew{I_{\text{SRL}}}=\frac{\Gamma_R \Gamma_L}{\Gamma_R +\Gamma_L}$ 
and  zero frequency Fano factor is \cite{deJong1996}
\begin{align}
 F_{\text{SRL}}(0)=\frac{\Gamma_R^2+ \Gamma_L^2}{(\Gamma_R +\Gamma_L)^2}.
\end{align}
Since the rates $\Gamma_L$ and $\Gamma_R$ are (positive) real numbers the Fano factor is always sub-Poissonian, $F(0)<1$.
The corresponding 
$g^{(2)}_{\text{SRL}}(\tau)$-function is given by 
\begin{align}
 g^{(2)}_{\text{SRL}}(\tau)=1-e^{-(\Gamma_R+\Gamma_L)\tau} \quad \text{and} \quad  g^{(2)}_{\text{SRL}}(0)=0.
\end{align}
Starting by zero at $\tau=0$ the $g^{(2)}_{\text{SRL}}$-function increases monotonously to one (not plotted), which indicates strictly anti-bunched electrons. 
Since, here, the Fano-factor is sub-Poissonian and the electron flow anti-bunched, the SRL model reflects the naive interpretation that sub-Poissonian statistics corresponds to anti-bunching.  Our second example  illustrates that this is not necessarily the case, however.

\subsection{Dynamical Channel Blockade}
%%%%%%%%%%%%%%%%%%%%%%%%%%%%%%%%%%%%%%%%%%%%%%%%%%%%%%%%%%%%%%%%%
\begin{figure}[tb]
  \begin{center}
\includegraphics[width=\columnwidth,clip]{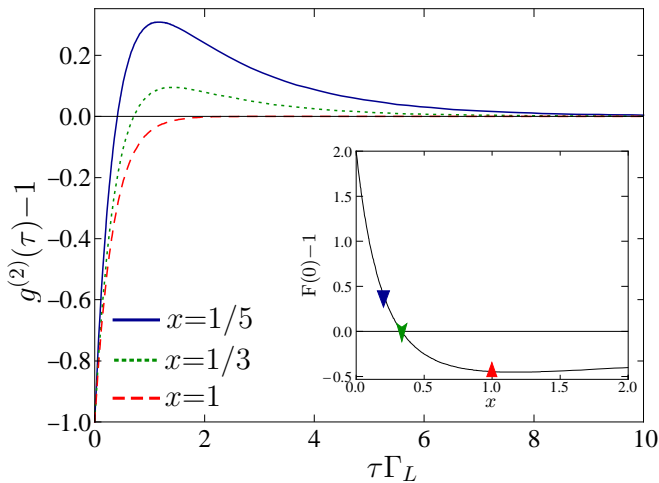}
   \caption{
    {\bf Main panel:} The second-order correlation function $g^{(2)}_{\text{DCB}}(\tau)-1$ for the Dynamical Channel Blockade model shown for three values of parameter $x=1,1/3,1/5$ corresponding to a sub-Poissonian (dashed), Poissonian (dotted) and super-Poissonian (line) Fano factor.
    {\bf Inset:} The (shifted) Fano factor $F_{\text{DCB}}(0)-1$ for the same model as a function of $x$.  The triangles mark the points for which the $g^{(2)}_{\text{DCB}}(\tau)$ is shown in the main panel.  
    Other parameters were: $\Gamma_L= \Gamma_R$.
     \label{DCB}} \end{center}
\end{figure}
%%%%%%%%%%%%%%%%%%%%%%%%%%%%%%%%%%%%%%%%%%%%%%%%%%%%%%%%%%%%%%%%%%%%%%%%
The dynamical channel blockade (DCB) model \cite{DCB-Belzig} is an important model of how interaction effects can give rise to super-Poissonian statistics. 
In its simplest version, the model consists of an empty state and a spin-up and a spin-down level, see \fig{models}(a).  In the basis $\bld{\rho}=\{\rho_0,\rho_\downarrow,\rho _\uparrow\}$, the Liouvillian reads
\begin{align}
  \mathcal{W}^\mathrm{DCB}(\chi) = 
  \rb{
    \begin{array}{ccc}
     - 2\Gamma_L 
         & \Gamma_R e^{i \chi}  & x \Gamma_R e^{i \chi}\\
      \Gamma_L  & -\Gamma_R  & 0 \\
   \Gamma_L  &  0  & -x \Gamma_R \\
    \end{array}    
  }
  \label{QD1_DCB2_kernel}
  .
\end{align}
The dimensionless factor $x$ changes the rate for outgoing spin-down electrons relative to spin-up ones. 
The steady-state current of the DCB model is
$
 \ew{I_{\text{DCB}}}=\frac{\Gamma_R \Gamma_L}{x(\Gamma_R +\Gamma_L)+\Gamma_L}
$,
and the zero-frequency Fano factor is \cite{DCB-Belzig}
\begin{align}
 F_{\text{DCB}}(0)=\frac{\Gamma_R^2 x+ \Gamma_L^2(3-2x+3x^2)}{(x\Gamma_R +x\Gamma_L+\Gamma_L)^2}.
\end{align}
For $x \mathord< \frac{2 \Gamma _L+\Gamma _R-\sqrt{\Gamma _R} \sqrt{8 \Gamma _L+\Gamma _R}}{2 \left(\Gamma _L-\Gamma_R\right)}$ ($x\mathord<\frac{1}{3}$ in the limit $\Gamma_L\mathord=\Gamma_R$), the Fano factor of this model is super-Poissonian, see \fig{DCB} (Inset), corresponding to DCB.

Our $g^{(2)}_{\text{DCB}}(\tau)$-function for this model reads.
\begin{align}
  g^{(2)}_{\text{DCB}}(\tau)=&1-  \frac{1}{2 x \gamma}\Big[\left(\gamma_L+x\gamma  \right) e^{-\frac{1}{2} \tau (2 \Gamma_L+x \Gamma_R+\Gamma_R+\gamma)} \nonumber \\ &-\left(\gamma_L-x\gamma  \right)
  e^{-\frac{1}{2} \tau ( +2 \Gamma_L+x \Gamma_R+\Gamma_R-\gamma)} \Big]
  ,
\end{align}
where $\gamma\mathord=\sqrt{4 \Gamma_L^2+(x-1)^2 \Gamma_R^2}$ and $\gamma_L=\Gamma_L(
   x^2+1)$.
\fig{DCB} shows ($g^{(2)}_{\text{DCB}}(\tau)-1$) plotted for $x\mathord=1,1/3$, and $1/5$ for which the corresponding Fano factors are 
$F_{\text{DCB}}(0)\mathord=0.55$, 
$F_{\text{DCB}}(0)\mathord=1$, and $F_{\text{DCB}}(0)\mathord=1.41$, respectively.

The sub-Poissonian case (represented by $x\mathord=1$ here) shows a monotonously increasing $g^{(2)}_{\text{DCB}}(\tau)$-function with interpretation identical to that of the SRL. 
In the $x=1/5$ case, the Fano factor is super-Poissonian and the function $g^{(2)}_{\text{DCB}}(\tau)-1$ starts off negative for $\tau=0$, but rises rapidly through zero and remains positive for large $\tau$.  For all times, $g^{(2)}_{\text{DCB}}(\tau) > g^{(2)}_{\text{DCB}}(0)$, and the electron flow is anti-bunched.
Thus, at least from the perspective of definition \eq{BvsAB}, we find contradiction 
in associating a super-Poissonian Fano factor with anti-bunching, as the flow here is anti-bunched independent of the Fano factor.
Nevertheless, there is a clear difference between super- and sub-Poissonian cases, since the $g^{(2)}$-function of the former exhibits a clear maximum and a negative slope for large times. The area of the $g^{(2)}$ curve then integrates to a positive number and the Fano factor is greater than one. 

This discrepancy can be arises because there are two competing processes occuring.  At short times the electrons must naturally be anti-bunched due to the strong CB.  At longer times, the DCB effect means however, that after one electron has tunnelled through the device, there is an increased probability that further electrons will tunnel once the system has had a chance to ``recharge'', and the electrons are indeed bunched, but only over time-scales larger than is required for the system recharge.  This analysis mirrors that of Kie\ss lich et al.
 \cite{KSW04} who studied the current-current correlation function for this model and obtained a similar result to \fig{DCB}.

The dotted-curve in \fig{DCB} shows the crossover case (here at $x=1/3$) where the Fano factor is exactly $F_{\text{DCB}}(0)=1$.  Here the  $g^{(2)}_{\text{DCB}}(\tau)$-function profile is qualitatively similar to the super-Poissonian case, but the ``negative-area'' exactly cancels the ``positive area'' resulting in $F_{\text{DCB}}(0)=1$.  From the point-of-view of the $g^{(2)}$-correlation function nothing particularly special occurs at a value $x=1/3$, although from the Fano factor, this may appear to be the case.  Note that, although at this point  $F_{\text{DCB}}(0)=1$, the complete statistics of the charge transfer are clearly not Poissonian since $g^{(2)}_{\text{DCB}}(\tau)\neq1$  is not equal to one at all times $\tau$.  This distinction can also be shown by calculating the higher-order Fano factors \cite{foot1}, which for this model deviate from the Poissonian value of one.

%%%%%%%%%%%%%%%%%%%%%%%%%%%%%%%%
\subsection{Double quantum dot}
\begin{figure}[t]
  \begin{center}
\includegraphics[width=\columnwidth]{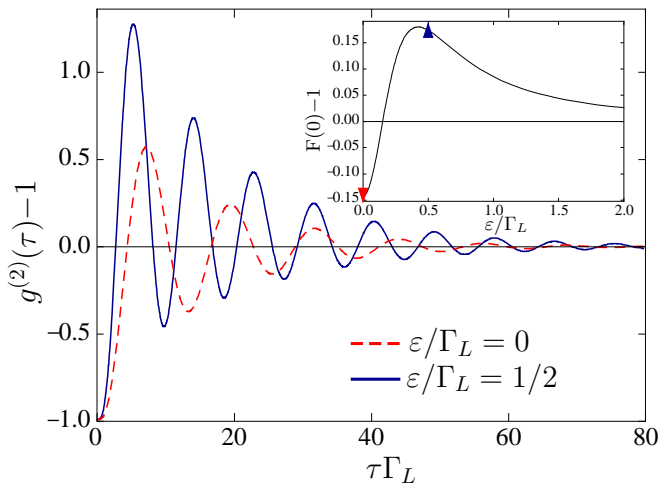}
   
  \caption{
    {\bf Main panel:} The $g^{(2)}_{\text{DQD}}(\tau)$ function for the double quantum dot (DQD) model with zero detuning such that the Fano factor is sub-Poissonian, $F_{\text{DQD}}(0)\approx0.859$ (dashed line); and detuning  $\varepsilon/\Gamma_L=1/2$ such that the Fano factor is super-Poissonian, $F_{\text{DQD}}(0)\approx1.174$ (solid line). 
    {\bf Inset:} The DQD Fano factor as a function of the detuning $\varepsilon$.The triangles mark the points for which the $g^{(2)}_{\text{DQD}}(\tau)\mathord-1$ is shown in the main panel.  
    Other parameters: $T_C/\Gamma_L =1/4$, $\Gamma_R/\Gamma_L=1/10$.
\label{DQD-sk}} 
\end{center}
\end{figure}

Our third example is the double quantum dot (DQD) in the strong CB, which allows us to study the influence of internal coherence on the $g^{(2)}$-function.  
A sketch of the DQD is shown in \fig{models}(b) and the system can either be empty or have a single electron in either of the left $\ket{L}$ or right  $\ket{R}$ states.
The DQD system Hamiltonian is then $H_{\text{DQD}}=\frac{\varepsilon}{2} (\op{L}{L}-\op{R}{R} )+T_C (\op{L}{R}-\op{R}{L} )$ with detuning $\varepsilon$ and coupling $T_C$. The left dot is coupled incoherently to the source lead and the right dot to the collector. With the system density matrix represented as $\bld{\rho}=\{\rho_{00},\rho_{LL},\rho_{RR},\rho_{LR},\rho_{RL}\}$,  the
system Liouvillian \cite{Gur,DQD1,Gur-F} reads
\begin{align}
\mathcal W^{\text{DQD}}(\chi) = \hspace{5.7cm} \nonumber \\
\begin{pmatrix} -\Gamma_L & 0 & e^{i \chi} \Gamma_R  & 0 & 0 \\
                           \Gamma_L  & 0 & 0 & i T_C & -iT_C \\
                            0 & 0 &-\Gamma_R &-i T_C & i T_C \\
                            0 & iT_C & -iT_C &-i\varepsilon-\Gamma_R/2 &0 \\
                            0 & -iT_C & iT_C & 0 & i \varepsilon -\Gamma_R /2\end{pmatrix}
                            .
\end{align}
The stationary current and the Fano factor of this model are\cite{Gur-F}
\beq
  \ew{I_{\text{DQD}}}&=&\frac{4 T_C^2 \Gamma _L \Gamma _R}{4 \Gamma _L \varepsilon ^2+4 T_C^2 \Gamma _R+\Gamma _L \left(8 T_C^2+\Gamma
     _R^2\right)}
     ;
  \nonumber\\
  F_{\text{DQD}}&=&\frac{16 \left(4 \Gamma _L^2+\Gamma _R^2\right) T_C^4+8 \Gamma _L^2 \left(12 \varepsilon ^2-\Gamma _R^2\right) T_C^2}
  {\left(4 \Gamma _L \varepsilon ^2+\Gamma _L \Gamma _R^2+4
     T_C^2 \left(2 \Gamma _L+\Gamma _R\right)\right){}^2}
  \nonumber \\
  &&+
  \frac{
    \Gamma _L^2 \left(4 \varepsilon ^2+\Gamma _R^2\right){}^2}{\left(4 \Gamma _L \varepsilon ^2+\Gamma _L \Gamma _R^2+4
     T_C^2 \left(2 \Gamma _L+\Gamma _R\right)\right){}^2}
     .
\eeq
Once the right tunnel rate $\Gamma_R$ is much smaller than the left tunnel rate $\Gamma_L$ the Fano factor can become super-Poissonian for finite detunings,
see \fig{DQD-sk}(Inset).
For zero detuning the Fano factor is sub-Poissonian, but as detuning increases, the Fano factor becomes super-Poissonian around $\varepsilon/\Gamma_L\approx0.15$ and reaches a maximum around $\varepsilon/\Gamma_L\approx0.4$, before decaying towards unity for large detunings.

In \fig{DQD-sk} we plot the $g^{(2)}_{\text{DQD}}(\tau)$ function for two sets of parameters: one that gives a sub-Poissonian Fano factor (dashed line) and one that gives a super-Poissonian one (continuous line).  In both cases, coherent tunneling between the two dots imprints damped oscillations onto $g^{(2)}_{\text{DQD}}(\tau)$ at finite $\tau$, similar to as is found in the waiting times $w(\tau)$ for a double quantum dot \cite{waiting1}, or in resonance fluorescence in quantum optics. This is in contrast to the other 
systems without quantum coherences discussed in this paper, where $g^{(2)}(\tau)$ either monotonously increases towards one or has a single peak and then decays towards one for $\tau\rightarrow \infty$.  We therefore expect oscillations in $g^{(2)}(\tau)$ to be an interesting tool for an experimental detection of quantum coherences.

Note again, that, as for all systems in the ultra-strong CB regime, $g^{(2)}_{\text{DQD}}(0)=0$ and so the electron flow is anti-bunched, according to definition \eq{BvsAB}. Once again though, this simple assignment misses the full complexity of the situation.
Both sub- and super-Poissonian traces are qualitatively very similar, suggesting once again that the value of the Fano factor is itself, not particularly diagnostic of the transport mechanisms occurring in this model.

%%%%%%%%%%%%%%%%%%%%%%%%%%%%%%%%%%%%%%%%%%%%%%%%%%%%%%%%%%%%%%%%%

\subsection{Doubly-occupied Quantum dots}
%%%%%%%%%%%%%%%%%%%%%%%%%%%%%%%%%%%%%%%%%%%%%%%%%%%%%%%%%%%%%%%%%
\begin{figure}[t]
  \begin{center}
\includegraphics[width=\columnwidth]{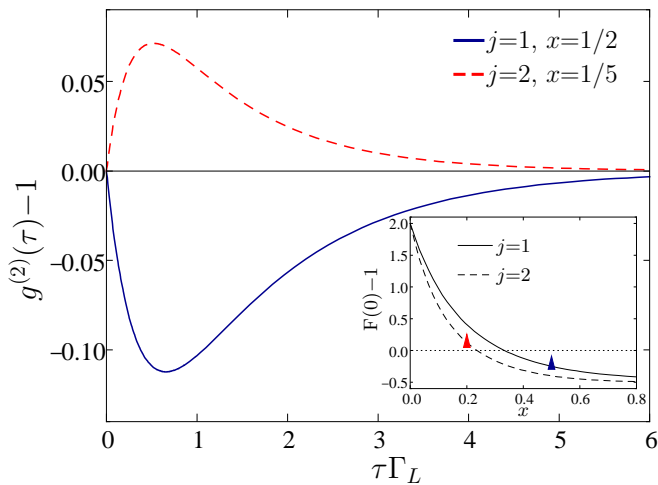}
 \caption{
    {\bf Main panel:}
    The $g^{(2)}_{2\text{el}}(\tau)$-function for the Anderson model with $j\in\left\{1,2\right\}$ single-electron channels.
    With a single channel ($j=1$) and rate parameter $x=1/2$ the Fano factor is sub-Poissonian ($F_{2\text{el}}(0)=3/4$, see inset), whereas the $g^{(2)}$-function clearly indicates bunching (solid line) for all times $\tau$.
    With two channels ($j=2$) and $x=1/5$ the Fano factor is super-Poissonian  ($F_{2\text{el}}(0)=9/8$) but the $g^{(2)}$-function indicates complete anti-bunching.
    This relationship between (anti-)bunching and Fano factor is completely opposite to the usual interpretation.
    {\bf Inset:} Fano factor $F(0)-1$ as a function of $x$ for $j=1$ (line) and $j=2$ (dashed). The triangles mark the $x$ for which $g^{(2)}_{\text{DQD}}(\tau)$ is plotted in the two cases where $F(0)\neq1$. Parameters: $\Gamma_L=\Gamma_R=\Gamma=1$. 
    \label{Two-el}
 }
  \end{center}
\end{figure}
%%%%%%%%%%%%%%%%%%%%%%%%%%%%%%%%%%%%%%%%%%%%%%%%%%%%%%%%%%%%%%%%%

Away from the strong CB limit, the system can be occupied by more than just a single electron at a given time. This opens the path for the $g^{(2)}$-function at $\tau=0$ to be non-zero, and the electron flow bunched.

As example of this class of model we consider a system with one empty state, $j\in \mathbb \{1,2\}$ single-occupied states (spin orbitals) and one double-occupied states within the transport window (and thus contributing to the current).  In the basis $\bld{\rho}=\{\rho_0,\rho_1=\sum_{j}\rho_{1(j)},\rho_{2}\}$, the kernel reads
\begin{align}
  \mathcal{W}^\mathrm{2\text{el}}(\chi) = 
  \rb{
    \begin{array}{ccc}
     - j\Gamma_L & \Gamma_R & 0\\
      j\Gamma_L &-\Gamma_L  -\Gamma_R & x j\Gamma_R e^{i \chi}  \\
     0&  \Gamma_L & -x j\Gamma_R  e^{i \chi} \\
    \end{array}    
  }
  \label{2en_DCB2_kernel}
  ,
\end{align}
where $x$ is a factor which modifies the rate of electrons tunneling out of the double occupied state.
For $j=2$ this model corresponds to the Anderson Model where both spin-up- and spin-down-levels are filled (emptied) with the same rate $\Gamma_L$ ($x \Gamma_R$). 
In contrast, for $j=1$, one of the singly-occupied levels lies {\em above} the transport window.  This can be realised, e.g.,  with a negative charging energy \cite{negU} and a magnetic field.
Figures \ref{models}(c) and \ref{models}(d) shows sketches of these two situations.

The steady-state current of this model with $\Gamma_L=\Gamma_R=\Gamma$ is given by
$
%\begin{align}
 \ew{I_{2\text{el}}}=\frac{2 j x \Gamma }{j x+x+1}
%\end{align}
$. 
The $g^{(2)}_{2\text{el}}(\tau)$-function is
\begin{align}
g^{(2)}_{\text{2el}}(\tau)=&1+\frac{ 1}{8 j \gamma_2 x} \Big[ (\text{a}_1 +\text{a}_2) e^{-\frac{1}{2} \tau (xj+j-\gamma_2+2) \Gamma }\nonumber \\
& +(\text{a}_1 -\text{a}_2) e^{-\frac{1}{2} \tau (x j+j+\gamma_2 +2) \Gamma } \Big],
\end{align}
with $\gamma_2= \sqrt{j^2 (x-1)^2+4} $, $\text{a}_1=2\gamma_2(1+x-3jx)$ and $\text{a}_2=-2 \big(2 (1 + x) + j^2 x (1 + x) + j (4 x-3  - 3 x^2)\big)$.

For $\tau=0$, we have
\begin{align}
  g^{(2)}_{2\text{el}}(0)=\frac{j x+x+1}{4 j x}, 
\end{align}
which is non-zero and becomes $g^{(2)}_{2\text{el}}(0)=1$ for $x=\frac{1}{3j-1}$. 
In \fig{Two-el} we plot ($g^{(2)}_{2\text{el}}(\tau)-1$) with this value of $x$ for $j=1$ (solid line) and $j=2$ (dashed line). 
Although both $g^{(2)}(\tau)\mathord-1$ curves  start at zero for $\tau=0$, the behaviour at finite time $\tau$ is completely different.  In the $j=1$ model, the $g^{(2)}_{2\text{el}}(\tau)$-function is negative and shows a minimum as a function of time $\tau$.  The electron flow is therefore bunched.   In the $j=2$ model, however, the $g^{(2)}_{2\text{el}}(\tau)$-function is everywhere positive, shows a maximum and indicates an anti-bunched electron flow. 
Looking at the the zero-frequency Fano factor (see \fig{Two-el}(Inset)) we see that it is {\em sub-Poissonian for the $j=1$ case} (negative area under $(g^{(2)}_{2\text{el}}(\tau)-1)$- function) and {\em super-Poissonian for $j=2$}  (positive area under $g^{(2)}_{2\text{el}}(\tau)-1$- function).  This is completely opposite to the intuitive understanding of the relationship between (anti-)bunching and the Fano factor.

%%%%%%%%%%%%%%%%%%%%%%%%%%%%%%%%%%%%%%%%%%%%%%%%%%%%%%%%%%%%%%
%%%%%%%%%%%%%%%%%%%%%%     END MATTER     %%%%%%%%%%%%%%%%%%%%
%%%%%%%%%%%%%%%%%%%%%%%%%%%%%%%%%%%%%%%%%%%%%%%%%%%%%%%%%%%%%%
\section{Conclusions}
The $g^{(2)}$- function is a well-known tool for  investigating bunching and anti-bunching behaviour of photon emission, but can equally well be used to describe the statistics of electron emission in transport systems such as quantum dots. The unifying frame in both cases is the concept of quantum jumps in Markovian quantum master equations that govern the dissipative dynamics. The relation \eq{Fano} between the $g^{(2)}$- function and the Fano factor clarifies that super-Poissonian statistics in electron transport do not necessarily correspond to bunching; sub-Poissonian statistics do not necessarily imply anti-bunching. 
As our examples show, single electron transport through nanostructures such as quantum dots offers the possibility to test these relations. 
An interesting further question would be the features of  cross correlations and  off-diagonal $g^{(2)}_{kl}$- functions defined with two different jump operators $\mathcal{J}_k\neq \mathcal{J}_l$.

\acknowledgments
We are grateful to Gerold Kie\ss lich for valuable discussions.
This work was supported by the DFG through GRK 1558.

%%%%%%%%%%%%%%%%%%%%%%%%%%%%%%%
\appendix
\section{A microscopic derivation of \eq{g2trans}}

We first specify the Hamiltonian of our transport system as composed of system, lead, and tunnel-coupling parts: $H=H_\mathrm{S}+H_\mathrm{leads}+H_\mathrm{T}$.
The system Hamiltonian we write as $H_\mathrm{S} = \sum_a E_a \op{a}{a}$, where $\ket{a}$ is a many-body system state of energy $E_a$.  We assume that the leads can be described by the non-interacting Hamiltonian
\beq
  H_\mathrm{leads} 
  = \sum_\alpha H^\alpha_\mathrm{leads} 
  =  \sum_\alpha\sum_{k} \omega_{\alpha k} c^\dagger_{\alpha k} c_{\alpha k}
  ,
\eeq
where $c_{\alpha k}$ is the annihilation operator for an electron in lead $\alpha$ with quantum numbers $k$ and energy $\omega_{\alpha k}$.
We assume that system and the leads are coupled with the single-particle tunnel Hamiltonian
\beq
  H_\mathrm{T} = \sum_{k,\alpha} 
  V_{k \alpha} c_{k\alpha}^\dagger d_\alpha
  +\mathrm{H.c.}
  ,
\eeq
where $d_\alpha$ is the annihilation operator of an electron in a localised system state, which we assume to be unique for each lead, and where $V_{k \alpha}$ is a tunnel amplitude.

To produce a correlation function analogous to \eq{g2full}, we first define for lead $\alpha$ the operator
\beq
  C^{(+)}_\alpha(t) =
  2i
  \sum_k V^*_{k \alpha} c_{k\alpha}(t) 
  \label{AP_bigC}
  .
\eeq
This definition is analogous to the expansion of electric field operator $E^{(+)}$ in terms of the normal-mode annihilation operators\cite {Loudon-book,ScullyBook}.  Here, the choice of coefficients is somewhat arbitrary, but $2 i V^*_{k \alpha}$ is convenient.
We then define our $g^{(2)}$-function for lead-$\alpha$ as
\beq
  g_\alpha^{(2)}(t,\tau) =
  \frac{
  \ew{
    C_\alpha^{(-)}(t) C_\alpha^{(-)}(t + \tau) C_\alpha^{(+)}(t+\tau) C_\alpha^{(+)}(t)
  }
  }
  {
  \ew{ C_\alpha^{(-)}(t) C_\alpha^{(+) }(t)}
  \ew{ C_\alpha^{(-)}(t+\tau) C_\alpha^{(+) }(t+\tau)}
  }
  .
  \nonumber\\
  \label{AP_g2transCdef}
\eeq

Heisenberg's equation of motion for the lead annihilators reads 
\beq
  \dot{c}_{k\alpha}(t) = 
  -i \omega_{k \alpha} c_{k\alpha} (t)
  - i V_{k\alpha} d_\alpha (t)
  .
\eeq
Introducing the Laplace transform $f(z) = \int_0^\infty dt e^{-z t} f(t)$ and solving gives
\beq
   c_{k\alpha}(z) = \frac{1}{z + i \omega_{k \alpha}}
   \left\{ 
     c_{k\alpha}(t=0) -i V_{k\alpha} d_\alpha (z)  
   \right\}
   \nonumber
\eeq
Summing over $k$ and regularizing  we obtain
\beq
  C^{(+)}_\alpha(z)
  &=&  
  \sum_k 
  2i V^*_{k\alpha}
  \left\{ 
     \pi \delta(i z - \omega_{k\alpha})
     +
   \frac{\mathds{P} }{z + i\omega_{k\alpha}}
   \right\}
  \nonumber\\
  && 
  ~~~~~~~~
  \times
  \left\{ 
     c_{k\alpha}(t=0) -i V_{k\alpha} d_\alpha (z)  
   \right\}
   ,
\eeq
where $\mathds{P}$ is the principal part.  We choose the initial state of lead $\alpha$ such that $c_{k\alpha}(t=0)$ evaluates to zero in the expectation value (see below).
Then
\beq
  C^{(+)}_\alpha(z)
  &=& 
  \left\{
    2 \pi  
    \sum_k 
    |V_{k\alpha}|^2\delta(i z - \omega_{k\alpha})
  \right.
  \nonumber\\
  && 
  ~~~ 
  \left.
    + 2i \mathds{P}
    \sum_k  |V_{k\alpha}|^2
    \frac{1}{i z -\omega_{k\alpha}}
  \right\}
  d_\alpha (z)  
  \nonumber\\
  &=&
  \left\{
    \Gamma_\alpha(iz) + i p_\alpha(iz)
  \right\}
  d_\alpha (z) 
\eeq
which defines the frequency-dependent rate,
$
  \Gamma(\omega) = 2 \pi \sum_k |V_{k\alpha}|^2
   \delta(\omega-\omega_{k\alpha})
$,
and principal part $p_\alpha(\omega)$.
If we assume that the rate is constant as a function of its argument, $\Gamma_\alpha(\omega) = \Gamma_\alpha$, and that the principal part vanishes, we obtain the result
\beq
  C^{(+)}(t) &=& 
  \Gamma_\alpha  d_\alpha (t) 
  \label{AP_Ceqd}
\eeq
These assumptions can be justified by assuming that lead $\alpha$ starts in the equillibrium state $\rho_\alpha^\mathrm{eq} = \exp\left\{-\beta (H^\alpha_\mathrm{leads}-\mu_\alpha N_\alpha)\right\}$ in the infinite-bias limit $\mu_\alpha \to - \infty$.  This also enforces the requirement $c_{k\alpha}(t=0) \rho_\alpha^\mathrm{eq} =0$.  

Placing result \eq{AP_Ceqd} in definition \eq{AP_g2transCdef} and taking the long-time limit yields 
\begin{align}
  g^{(2)}_\alpha(\tau) =
  \frac{
    \ew{
      d^\dag_\alpha
      d^\dag_\alpha(\tau) 
      d_\alpha(\tau) 
      d_\alpha
    }_\mathrm{stat}
  }
  {
    \ew{
      d^\dag_\alpha
      d_\alpha
    }_\mathrm{stat}^2
  }
  %;\quad 
  %\tau >0
  ,
\end{align}
analogous to \eq{g2sigmas}.

If we assume that all leads start in this same infinite-bias limit, the time-evolution of the system density matrix can be described by the Markovian master equation\cite{FCS3}:
\beq
  \dot \rho 
  %&=& \mathcal{W}  \rho
  %\nonumber\\
  &=&
  -i \left[H_\mathrm{S}, \rho\right]
  +
  \sum_{\alpha \in S} \Gamma_{\alpha}
  \left\{
    d^\dag_\alpha \rho d_\alpha
    - \frac{1}{2}\rho  d_\alpha d^\dag_\alpha 
    - \frac{1}{2} d_\alpha d^\dag_\alpha\rho
  \right\}
  \nonumber\\
  &&
  +
  \sum_{\alpha \in D} \Gamma_{\alpha}
  \left\{
    d_\alpha \rho d^\dag_\alpha
    - \frac{1}{2}\rho  d^\dag_\alpha d_\alpha 
    - \frac{1}{2} d^\dag_\alpha d_\alpha\rho
  \right\}
  \label{LPTinfinitebias}
  ,
\eeq
where the first sum is over source leads, the second over drains and where $\Gamma_\alpha$ are rates.
Writing this as $\dot \rho = ( \mathcal{W}_0 + \mathcal{J})\rho$ with jump super-operator  $\mathcal{J}\rho = \Gamma_\alpha d_\alpha \rho d^\dag_\alpha$, we see that definitions \eq{AP_bigC} and \eq{AP_g2transCdef} indeed reproduce \eq{g2trans} of the main text.

%%%%%%%%%%%%%%%%%%%%%%%%%%%%%%%%%%%%%%%%%%%%%%%%%%%%%%%%%%%%%%%%%%%%%

%%%%%%%%%%%%%%%%%%%%%%%%%%%%%%%%%%%%%%%%%%%%%%%%%%%%%%%%%%%%%%%%%%%%%%%%%

\begin{thebibliography}{99}
\bibitem{foot2} We refer to the definition of $Q=R\theta(T)$  in \cite{Mandel-Wolf}, Ch. 14.9.2 with $\lim_{T\to\infty} \theta(T) = 2\int_0^\infty d\tau \lambda(\tau)$ (ib., Ch. 15.6.6)
and the identification of the counting rate $R= \eww{\mathcal{J}}$. With $\lambda(\tau)$ corresponding to our $S(\tau)/  \eww{\mathcal{J}}^2 =g^{(2)}(\tau)-1$ we obtain
$F(0)=Q+1 = 1+ 2  \eww{\mathcal{J}} \int_0^\infty d\tau \lambda(\tau)$, thus \eq{Fano}.
\bibitem{foot1}Here we always calculate the second order Fano factor hence the second current cumulant normalized by the current. The $n$-th order Fano factor defined analogues as $n$-th order current correlation functions, normalized by the stationary current. \cite{FCS2,FCS3}
\end{thebibliography}
\end{document}